\documentclass[twocolumn,superscriptaddress,PRB]{revtex4-1}

\usepackage{color}

\usepackage{amsmath}
\usepackage{amssymb}
\usepackage{graphicx}
\usepackage{bm}
\usepackage[colorlinks,linkcolor=blue, urlcolor=blue,citecolor=blue,bookmarks=fales]{hyperref}

\begin{document}
\title{Zero-energy vortex bound state in the superconducting topological surface state of Fe(Se,Te)}

\author{T. Machida}
\email{tadashi.machida@riken.jp}
\affiliation{RIKEN Center for Emergent Matter Science, 2-1 Hirosawa, Wako, Saitama 351-0198, Japan}

\author{Y. Sun}
\affiliation{Department of Physics and Mathematics, Aoyama Gakuin University, 5-10-1 Fuchinobe, Chuou-ku, Sagamihara, Kanagawa 252-5258, Japan}

\author{S. Pyon}
\affiliation{Department of Applied Physics, University of Tokyo, 7-3-1 Hongo, Bunkyo-ku, Tokyo 113-8656, Japan}

\author{S. Takeda}
\affiliation{Laboratory for Materials and Structures, Tokyo Institute of Technology, 4259 Nagatsuda, Midori-ku, Yokohama, Kanagawa 226-8503, Japan}

\author{Y. Kohsaka}
\affiliation{RIKEN Center for Emergent Matter Science, 2-1 Hirosawa, Wako, Saitama 351-0198, Japan}

\author{T. Hanaguri}
\email{hanaguri@riken.jp}
\affiliation{RIKEN Center for Emergent Matter Science, 2-1 Hirosawa, Wako, Saitama 351-0198, Japan}

\author{T. Sasagawa}
\affiliation{Laboratory for Materials and Structures, Tokyo Institute of Technology, 4259 Nagatsuda, Midori-ku, Yokohama, Kanagawa 226-8503, Japan}

\author{T. Tamegai}
\affiliation{Department of Applied Physics, University of Tokyo, 7-3-1 Hongo, Bunkyo-ku, Tokyo 113-8656, Japan}

\maketitle
 \textbf{
Majorana quasiparticles (MQPs) in condensed matter play an important role in strategies for topological quantum computing~\cite{Nayak_RMP_2008,Alicea_RPP_2012,Beenakker_ARCMP_2013,Elliott_RMP_2015,Sato_RPP_2017} but still remain elusive.
Vortex cores of topological superconductors may accommodate MQPs that appear as the zero-energy vortex bound state (ZVBS)~\cite{Ivanov_PRL_2001,Fu_PRL_2008}.
An iron-based superconductor Fe(Se,Te) possesses a superconducting topological surface state~\cite{Wang_PRB_2015,Wu_PRB_2016,Xu_PRL_2016,Zhang_Science_2018} that has been investigated by scanning tunneling microscopies to detect the ZVBS~\cite{Wang_Science_2018,Chen_NC_2018}.
However, the results are still controversial~\cite{Wang_Science_2018,Chen_NC_2018}.
Here, we performed spectroscopic-imaging scanning tunneling microscopy with unprecedentedly high energy resolution to clarify the nature of the vortex bound states in Fe(Se,Te).
We found the ZVBS at 0$\bm{\pm}$20~$\bm{\mu}$eV suggesting its MQP origin, and revealed that some vortices host the ZVBS while others do not.
The fraction of vortices hosting the ZVBS decreases with increasing magnetic field, while chemical and electronic quenched disorders are apparently unrelated to the ZVBS.
These observations elucidate the conditions to achieve the ZVBS, and may lead to controlling MQPs.
}
\\

The Majorana fermion is an exotic particle that is its own antiparticle.
Although it is not established whether or not the Majorana fermion exists in nature as an elementary particle, it can emerge in condensed matter systems as MQPs.
MQPs are potentially capable of future applications such as fault-tolerant quantum computing based on their adiabatic braiding~\cite{Nayak_RMP_2008,Alicea_RPP_2012,Beenakker_ARCMP_2013,Elliott_RMP_2015,Sato_RPP_2017,Ivanov_PRL_2001,Fu_PRL_2008}.

Although several ways have been proposed to achieve MQPs, their implementations in real systems are challenging.
Many of the efforts rely on the boundary states of topological $p$-wave superconductors that could be realized in spin-polarized low-dimensional systems.
These include one-dimensional Rashba semiconductor nanowires~\cite{Mourik_Science_2012} and magnetic-atom chains~\cite{Nadj-Perge_Science_2014,Kim_SciAdv_2018} in proximity to $s$-wave superconductors.
In two-dimensional systems, the interface between the spin-polarized surface state of a topological insulator and an $s$-wave superconductor can exhibit topological $p$-wave superconductivity, and accordingly, vortex cores may accommodate MQPs, which result in the ZVBS~\cite{Fu_PRL_2008}.

The above platforms all require artificial heterostructures that demand complicated sample preparation techniques.
Recently, bulk superconductors with spin-polarized topological surface states attract much attention because topological $p$-wave superconductivity may be naturally induced at the surfaces~\cite{Hosur_PRL_2011}.
These materials not only simplify sample preparations but also make spectroscopic-imaging (SI) scanning tunneling microscopy (STM) experiment easier because MQPs in the vortex cores are not buried in the interface but exposed at the surface.
Among various candidate materials~\cite{Guan_SciAdv_2016,Iwaya_NatCommun_2017}, an iron-based superconductor Fe(Se,Te) is promising~\cite{Wang_PRB_2015,Wu_PRB_2016,Xu_PRL_2016,Zhang_Science_2018}.
While FeSe has a topologically-trivial band structure, Te substitution introduces stronger spin-orbit interaction and promotes larger overlapping between the chalcogen $p_z$ orbitals, leading to the topologically non-trivial spin-polarized Dirac cone at the surface~\cite{Wang_PRB_2015,Wu_PRB_2016,Xu_PRL_2016}.
Recent angle- and spin-resolved photoemission spectroscopy has confirmed such a Dirac cone~\cite{Zhang_Science_2018}.

Besides this topological property, Fe(Se,Te) is unique because its superconducting gap $\Delta$ is large ($\sim 1.5$~meV~\cite{Hanaguri_Science_2010}) whereas its Fermi energy $E_{\mathrm F}$ is very small ($\sim 10$~meV~\cite{Rinott_SciAdv_2017}).
This makes easier to distinguish the ZVBS from the conventional Caroli-de Gennes-Matricon (CdGM) bound states in the vortex core~\cite{Caroli_PL_1964}.
The energies of the CdGM states are given by $\sim \mu \Delta^{2}/E_{\mathrm F}$, where $\mu$ is a half integer.
In principle, CdGM states appear only at finite energies and are thus distinguishable from the ZVBS, but the level splitting in most superconductors is much smaller than the various energy-broadening factors in actual experiments.
As a result, many CdGM states overlap each other, forming a broad zero-energy peak in the quasiparticle-excitation spectrum.
Thus it is generally tricky to establish the relation between the ZVBS and the MQP~\cite{Xu_PRL_2015,Sun_PRL_2016}.
The situation becomes simpler in Fe(Se,Te) because the lowest CdGM-state energy is expected to be high ($\sim 100$~$\mu$eV), being distinguishable from the ZVBS by currently available SI-STM technology.

\begin{figure*}[t]
\includegraphics[width=16.5cm]{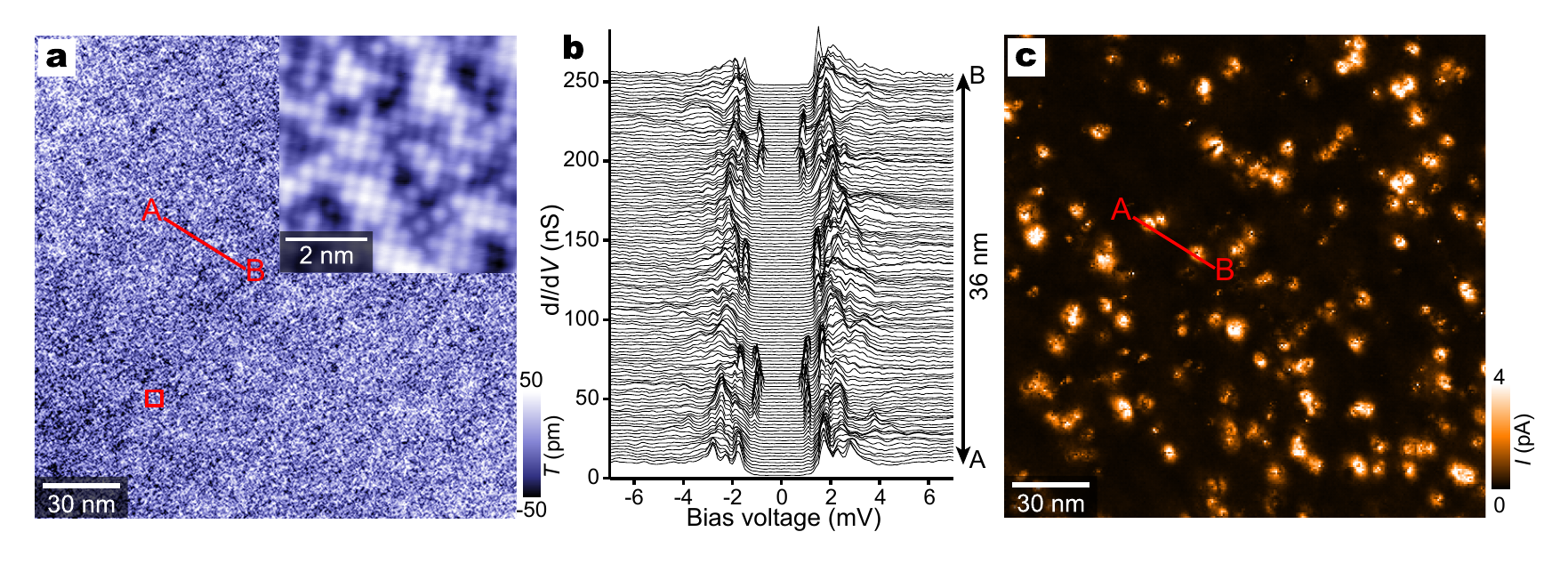}
\caption{\label{Fig1}
\textbf{Quenched disorders in Fe(Se,Te).}
\textbf{a}, A constant-current topographic image taken over a FOV of 187~nm~$\times$~187~nm.
The set-point of the feedback was a tunneling current of $I = 100$~pA at sample bias voltage $V = -40$~mV.
Inset depicts a magnified image of the area shown by the red box in the main panel.
\textbf{b}, A line profile of tunneling spectra taken along the 36~nm-long path shown by a red line in \textbf{a}.
The bias modulation amplitude $V_\mathrm{mod}$ was set to 70.7~$\mu$V$_{\rm{rms}}$ with the set-point of $I = 100$~pA at $V = -10$~mV.
\textbf{c}, A current map $I(\textbf{r}, E~=~1~\mathrm{meV})$ taken in the same FOV as \textbf{a}.
The set-point was $I = 100$~pA at $V = -10$~mV.
All data were taken at $T_{\rm{eff}} \sim 85$~mK.
}
\end{figure*}

SI-STM searches for the ZVBS in Fe(Se,Te) have been performed by two groups; one group indeed observed the ZVBS~\cite{Wang_Science_2018}, whereas the other reported its absence~\cite{Chen_NC_2018}.
Besides this apparent contradiction, both experiments have been done at temperatures $T \sim 0.5$~K with energy resolutions of $\sim 250$~$\mu$eV, which is comparable to or larger than the expected lowest CdGM state energy.
Therefore, a more comprehensive SI-STM experiment with higher energy resolution is indispensable to make clear the nature of the ZVBS in Fe(Se,Te) and its relation to MQPs.
Here we have inspected a large number of vortices at different magnetic fields at an effective electron temperature $T_{\rm{eff}} \sim 85$~mK using a dilution refrigerator STM system~\cite{Machida_RSI_2018} with unprecedentedly high energy resolution of $\sim 20$~$\mu$eV.

We first characterized the sample in zero magnetic field and found various quenched disorders.
Figure~1\textbf{a} shows an STM topograph of the field of view (FOV) where the following SI-STM experiments have been performed.
A magnified topograph (Fig.~1\textbf{a} inset) depicts a square lattice of the topmost chalcogen atoms with randomly distributed depressions that correspond to the Se atoms whose atomic radius is smaller than that of Te.
We identified locations of all Se atoms over the FOV, enabling us to quantify the spatial variation of the local Se density as well as to determine the overall chemical composition to be FeSe$_{0.4}$Te$_{0.6}$.
(See Supplementary Information Section I for the estimation of Se density).
Figure~1\textbf{b} displays position $\mathbf{r}$ dependence of the tunneling conductance spectrum $g(\mathbf{r}, E=eV) \equiv dI/dV$, which reflects local density of states (LDOS), taken along the line shown in Fig.~1\textbf{a}.
Here, $E$ is the electron energy, $e$ is the elementary charge, $I$ is the tunneling current and $V$ is the sample bias voltage.
The spectral weight is missing over an extended energy range $|E| \lesssim 1.5$~meV, indicating fully-gapped superconductivity~\cite{Hanaguri_Science_2010,Massee_SciAdv_2015,Yin_NP_2015}.
The spectrum near the gap-edge energy consists of multiple LDOS peaks with pronounced heterogeneity.
At some locations, there appear in-gap bound states that indicate defects acting as pair breakers.
Figure~1\textbf{c} shows spatial distribution of such defects obtained by mapping $I(\mathbf{r}, E=1\mathrm{~meV})$, which reflects integrated LDOS within the superconducting gap (Fig.~1\textbf{c}).
(See Supplementary Information Section II for detailed defect states.)

\begin{figure*}[t]
\includegraphics[width=16.5cm]{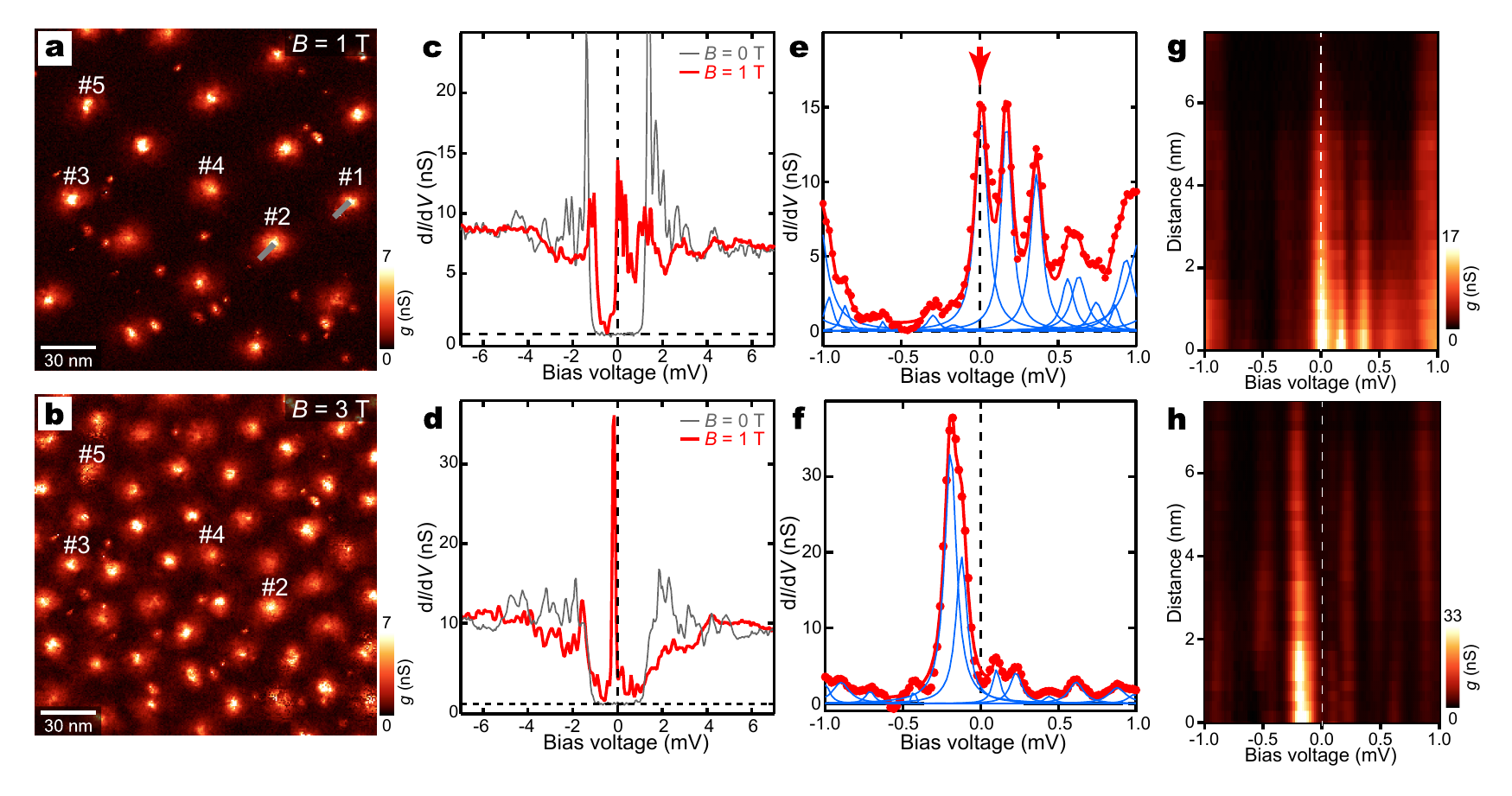}
\caption{\label{Fig2}
\textbf{Vortex-core dependent bound states.}
\textbf{a},\textbf{b}, Zero-energy conductance maps in the same FOV as Fig.~1\textbf{a} at $B = 1$ and 3~T, respectively.
Measurement conditions are $I = 100$~pA, $V = -10$~mV and $V_\mathrm{mod} = 354$~$\mu$V$_{\mathrm{rms}}$.
The vortices identified as \#2 - \#5 reside at the same locations at $B$ = 1 and 3~T.
\textbf{c},\textbf{d}, Red curves represent tunneling spectra at the vortex centers of \#1 and \#2 in Fig. 2\textbf{a}, respectively.
Spectra taken at the same locations without a magnetic field are shown by gray curves.
Measurements were done with $I = 100$~pA, $V = -10$~mV and $V_\mathrm{mod} = 35.4$~$\mu$V$_{\mathrm{rms}}$.
\textbf{e},\textbf{f}, Detailed tunneling spectra at the vortex centers of \#1 and \#2, respectively.
The conductance data (red circles) were obtained by the numerical differentiation of the $I$-$V$ curves with the energy sampling interval of $20$~$\mu$eV.
The set-point was $I = 100$~pA and $V = -10$~mV.
Red curves denote the fitting results obtained using the multiple Lorentzian functions (blue lines).
The details of the fitting procedure are described in Supplementary Information Section III.
\textbf{g},\textbf{h}, Line profiles of the tunneling spectra at $B$ = 1 T for \#1 and \#2, respectively, along the gray lines in Fig. 2\textbf{a}.
Measurements were done at the same conditions as \textbf{e} and \textbf{f}.
Distances are measured from the vortex centers.
All data were taken at $T_{\rm{eff}} \sim 85$~mK.
}
\end{figure*}

Next we examined spectroscopic features of vortices in such an inhomogeneous environment.
Vortices are imaged as high $g(\mathbf{r}, E=0\mathrm{~meV})$ regions as shown in Fig.~2\textbf{a},\textbf{b} for magnetic fields $B=1$~T and 3~T, respectively.
First we focused on two vortices labeled as \#1 and \#2 in Fig.~2\textbf{a} and performed detailed tunneling spectroscopies at the vortex-center locations with and without $B$ (Fig.~2\textbf{c},\textbf{d}).
The vortex center was defined as the position where $g(\mathbf{r}, E=0\mathrm{~meV})$ takes its maximum in the vortex.
At both locations, vortices suppress the spectral weights near the gap edges and induce multiple vortex bound states inside of the superconducting gap.
We performed higher energy resolution ($\sim 20$~$\mu$eV) spectroscopy and fitted the obtained spectrum using multiple Lorentzian functions to determine the energies of the bound states.
(See Supplementary Information Section III for the multiple-peak fitting).
As shown in Fig.~2\textbf{e},\textbf{f}, vortex \#1 possesses the ZVBS, whereas \#2 does not.
The finite-energy bound states are also different between \#1 and \#2 but share similarities in that they appear in pairs at nearly symmetric energies with large asymmetry in their intensities.
These are reminiscent of the behavior of the CdGM states in the quantum-limit vortex~\cite{Chen_NC_2018}, whereas Friedel-like oscillations, which are another signature of the quantum-limit vortex~\cite{Hayashi_PRL_1998}, are not resolved in the spatial variation of the bound states (Fig.~2\textbf{g},\textbf{h}).

\begin{figure*}[t]
\includegraphics[width=16.5cm]{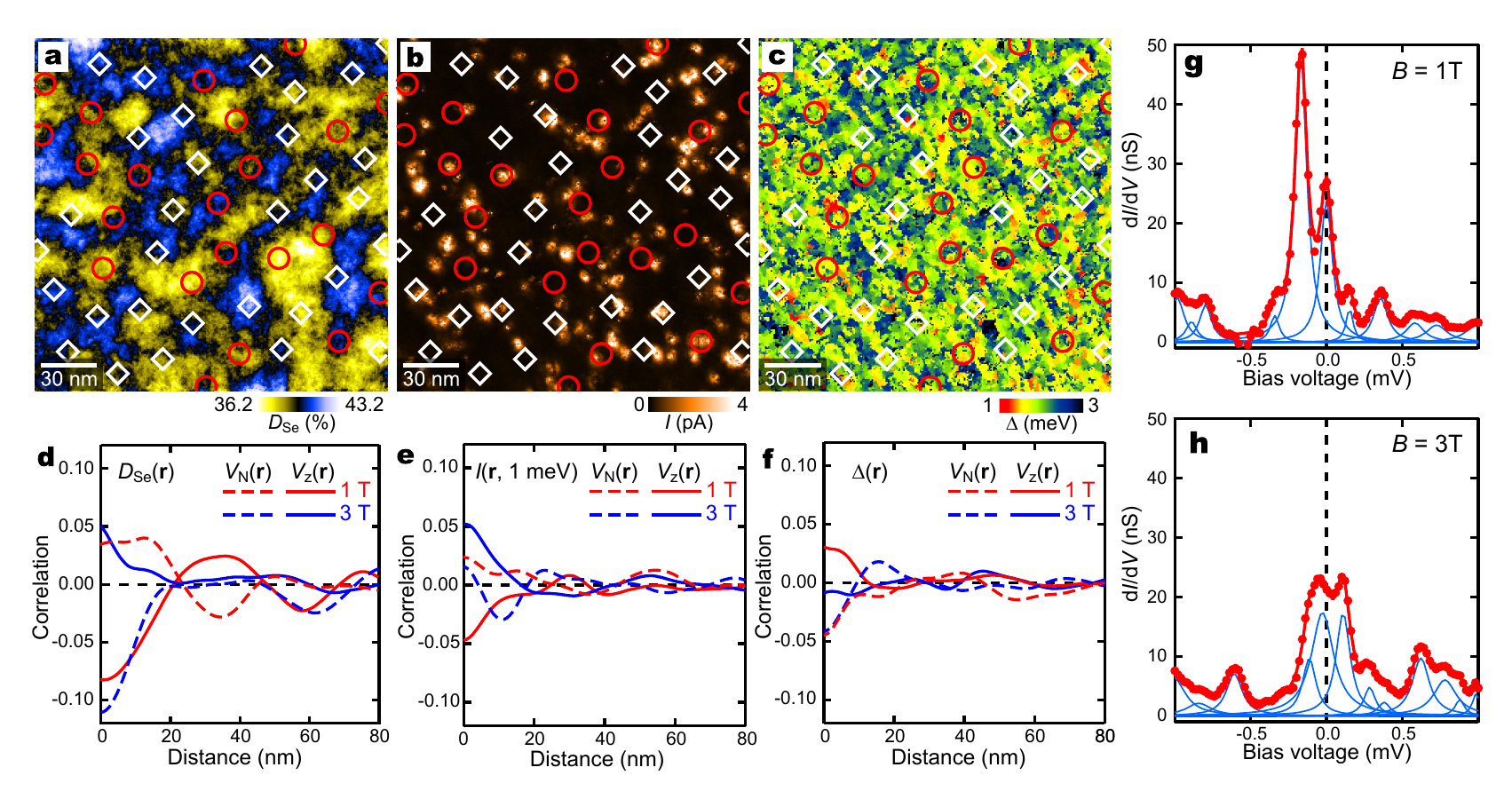}
\caption{\label{Fig3}
\textbf{Spatial correlations between the vortices and quenched disorders.}
\textbf{a}, A local Se density map $D_{\mathrm{Se}}(\mathbf{r})$ defined in a circle centered at the position $\mathbf{r}$ with a diameter of 20~nm.
(Detailed diameter dependence is described in Supplementary Information Section I.)
\textbf{b}, A current image $I(\mathbf{r}, E=1~\mathrm{meV})$ showing all the zero-field in-gap bound states.
\textbf{c}, A local gap-amplitude map $\Delta(\mathbf{r})$ defined as the energy at which the tunneling conductance takes its maximum on the positive-bias side.
Details are discussed in Supplementary Information Section IV.
For visual comparison between the ZVBS and the quenched disorders shown in \textbf{a}-\textbf{c}, the locations of the vortices with (red circles) and without (white diamonds) the ZVBS at $B = 3$~T are superimposed on each image.
\textbf{d}-\textbf{f}, Azimuthally-averaged cross-correlations of the simplified vortex images (see text for details) and quenched disorders shown in \textbf{a}-\textbf{c}, respectively.
Solid and dashed curves represent the correlations between the quenched disorders and the positions of vortices with and without ZVBS, respectively.
Red and blue curves are for $B = 1$ and 3~T, respectively.
\textbf{g},\textbf{h}, Tunneling spectra taken at the centers of vortices identified as ``\#3" in Fig. 2\textbf{a} ($B$ = 1~T) and Fig. 2\textbf{b} ($B$ = 3~T), respectively.
These vortices reside at the same location in both fields.
Solid red circles denote the experimental data and the red curves are the fitting results obtained using multiple Lorentzian functions shown in blue.
All data were taken at $T_{\rm{eff}} \sim 85$~mK.
}
\end{figure*}

\begin{figure*}[t]
\includegraphics[width=16.5cm]{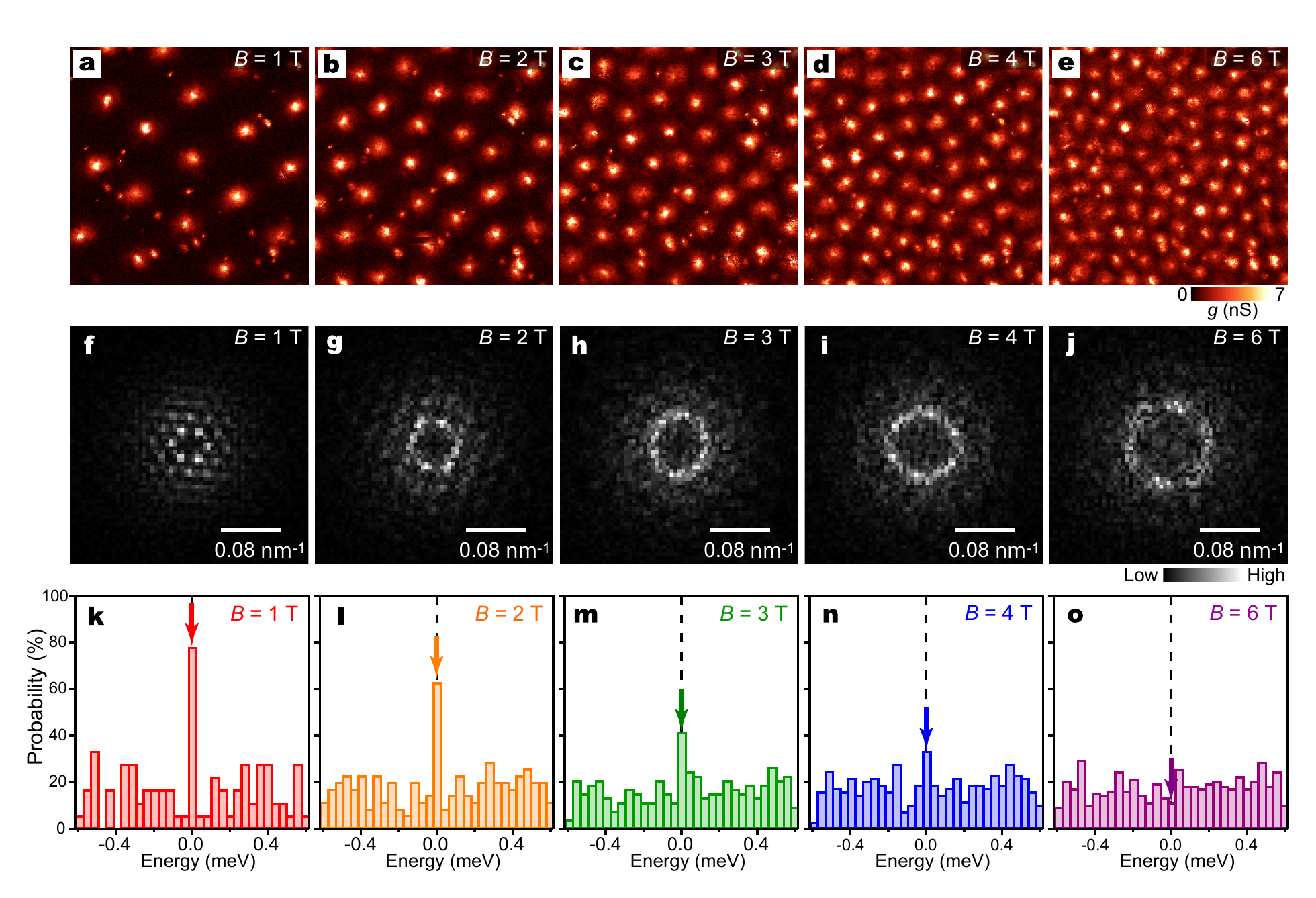}
\caption{\label{Fig4}
\textbf{Magnetic field dependence of the vortex lattice and the ZVBS.}
\textbf{a}-\textbf{e}, Images of vortices obtained by mapping $g(\mathbf{r}$, 0 meV) in the same FOV at $B$ = 1~T,  2~T, 3~T, 4~T, and 6~T, respectively.
Measurement conditions were $I = 100$~pA, $V = -10$~mV and $V_\mathrm{mod} = 354$~$\mu$V$_{\mathrm{rms}}$.
\textbf{f}-\textbf{j}, Fourier-transformed images of \textbf{a}-\textbf{e}, respectively.
\textbf{k}-\textbf{o}, Histograms of the appearance probability of the peaks in the tunneling spectra at given energies.
All the imaged vortices in this FOV were used for the analyses at all $B$.
All data were taken at $T_{\rm{eff}} \sim 85$~mK.
}
\end{figure*}

It is important to investigate whether or not the observed two types of vortices, with and without the ZVBS, are related to any of the quenched disorders in this material.
To this end, we measured high-resolution tunneling spectra at all of the vortex centers in Fig.~2\textbf{a},\textbf{b} and performed multiple Lorentzian fitting to identify vortices with the ZVBS.
We found that about 80~\% and 40~\% of vortices host the ZVBS (peak energy $|E| < 20$~$\mu$eV) at $B=1$~T and 3~T, respectively.
In Fig. 3\textbf{a}-\textbf{c}, the locations of vortices with (red circles) and without (white diamonds) the ZVBS at $B=3$~T are shown on the images of the Se-density map $D_{\mathrm{Se}}(\mathbf{r})$, defect-location map $I(\mathbf{r}, E=1\mathrm{~meV})$, and superconducting-gap map $\Delta(\mathbf{r})$, respectively.
(Procedures for making these images are described in Supplementary Information Section I, II, and IV, respectively.)
Apparently, neither the vortices with nor without the ZVBS correlate with any of the quenched disorders.

We carried out more quantitative analyses by calculating cross-correlation functions between the vortex images and the images of quenched disorders.
For the vortex images, each vortex is represented by the 2-dimensional Gaussian function that is common to all the vortices.
The resultant simplified vortex image is then separated into two images, $V_{\mathrm{Z}}(\mathbf{r})$ and $V_{\mathrm{N}}(\mathbf{r})$, which represent distributions of vortices with and without the ZVBS, respectively (See Supplementary Information Section V for details).
Figure~3\textbf{d}-\textbf{f} depicts azimuthally-averaged cross-correlation functions between $V_{\mathrm{Z}}(\mathbf{r})$ or $V_{\mathrm{N}}(\mathbf{r})$ and the images of quenched disorders.
In all the cases, correlations are small ($\lesssim 0.1$) and do not exhibit systematic trends, indicating that the observed quenched disorders are nothing to do with the ZVBS.

These observations suggest that the ZVBS is not governed by preexisting local chemical and electronic landscapes.
To confirm this conjecture, we searched for the vortices that share the same positions at $B=1$~T and 3~T.
We found four such vortices in our FOV (vortices labeled as \#2 to \#5 in Fig.~2\textbf{a},\textbf{b}).
Figure~3\textbf{g},\textbf{h} shows spectra taken at one of these vortices (\#3) at 1~T and 3~T, respectively.
The overall spectral shapes are clearly different between 1~T and 3~T.
(Spectra for other vortices are shown in Supplementary Information Section VI.)
This evidently indicates that the non-local effects of surrounding vortices and/or the field strength itself play an important role.

Given the observation that the fraction of vortices with the ZVBS is smaller at $B=3$~T than that at 1~T, we investigated $B$ dependence systematically.
Figure~4\textbf{a}-\textbf{e} shows a series of $g(\mathbf{r}, E=0~\mathrm{meV})$ images at different $B$ in the same FOV.
Increasing $B$ not only generates more vortices but also enhances $g(\mathbf{r}, E=0~\mathrm{meV})$ in between the vortices.
This means that vortices are overlapping each other at higher $B$.
Fourier-transformed images of $g(\mathbf{r}, E=0~\mathrm{meV})$ maps (Fig.~4\textbf{f}-\textbf{j}) show ring-like features especially at higher $B$ rather than the six-fold spots, indicating that the orientation correlation in the vortex lattice is lost whereas the distance correlation remains.
We examined the $B$ effect on the ZVBS by inspecting high resolution tunneling spectra at all vortex centers in Fig.~4\textbf{a}-\textbf{e} (222 in total).
At each $B$, we fitted all the spectra by multiple Lorentzian functions and made a histogram of the appearance probability of the bound-state peaks binned by the peak energy (Fig.~4\textbf{k}-\textbf{o}).
The appearance probabilities of the finite energy peaks are almost independent of $B$, whereas the probability of the ZVBS decreases with increasing $B$.
It should be noted that the probability of the ZVBS is much higher than that of the finite energy peaks, indicating that the ZVBS is distinct from other bound states.
The same features have been reproduced in a different FOV (See Supplementary Information Section VII).

Our experiments have revealed important aspects of the ZVBS in Fe(Se,Te).
First, we note that it is unlikely that the observed ZVBS is due to the lowest CdGM state at $\sim \Delta^{2}/2E_{\mathrm F}$.
We have observed the ZVBS within $|E|\leq 20$~$\mu$eV at the location where $\Delta(\mathbf{r}) \sim 1.5$~meV (Fig.~2\textbf{c}).
If this ZVBS could be the lowest CdGM state, $E_{\mathrm F}$ should be larger than $\sim 50$~meV, contradicting the angle-resolved photoemission spectroscopy results~\cite{Rinott_SciAdv_2017,Zhang_Science_2018}.
The MQP state is the prime candidate for an explanation of the ZVBS~\cite{Wang_Science_2018}.
In this scenario, the question is why the magnetic field suppresses the appearance probability of the ZVBS.
It has been pointed out that MQPs in a vortex lattice form a Majorana band through the interaction between vortices~\cite{Biswas_PRL_2013,Chiu_PRB_2015,Liu_PRB_2015}, and the ZVBS may split away from zero energy~\cite{Chiu_PRB_2015,Liu_PRB_2015}.
Indeed, enhanced $g(\mathbf{r}, E=0~\mathrm{meV})$ in between the vortices at higher $B$ (Fig.~4\textbf{a}-\textbf{e}) indicate that vortices are not independent of each other.
Meanwhile, if the vortex lattice is regular, all vortices must be the same.
The observed pronounced variations in the vortex spectra indicate that the effect of disorders should be taken into account.
Since the chemical and electronic quenched disorders are not correlated with the ZVBS, it is plausible that the observed coexistence of vortices with and without the ZVBS is related to disorder in the vortex arrangements.
Because the distance correlation of vortices remains, we speculate that the orientational disorder may affect the ZVBS in some form.
It is an important future issue to theoretically investigate the effect of vortex disorder on MQPs.
Another interesting theoretical challenge is the effect of magnetic-field strength, such as the Zeeman effect.
In any case, our experimental observations provide phenomenological conditions to achieve the ZVBS in Fe(Se,Te).
The apparent chemical and electronic disorders do not play a major role but it is crucial to keep the magnetic field low.
This is an important clue for the control of the MQP.

\subsection*{Method}
\textbf{Single crystal growth}

Single crystals were grown from the melt with nominal composition of FeSe$_{0.4}$Te$_{0.6}$ and were annealed in appropriate O$_{2}$ partial pressure at 400~$^{\circ}$C to remove the interstitial excess iron atoms.
Detailed crystal growth and annealing procedures have been described elsewhere~\cite{Sun_SST_2012,Sun_SRep_2014}.
The superconducting transition temperature of the samples used in this work is determined to be 14.5~K by magnetization measurements.

\textbf{SI-STM experiments}
We utilized a dilution-refrigerator-based ultra-high-vacuum (UHV) ultra-low-temperature STM~\cite{Machida_RSI_2018}.
Electrochemically etched tungsten wires were used for the scanning tips that were cleaned by Ar-ion sputtering and subsequent electron-beam heating in a UHV chamber.
The tips were further conditioned by controlled indentation into a clean Au(100) surface before loading the Fe(Se,Te) samples.
The clean surfaces needed for SI-STM were prepared by \textit{in situ} cleaving in UHV at liquid-nitrogen temperature.
Most of the spectroscopic measurements were done by using the standard lock-in method with a bias modulation frequency of 617.4~Hz.
For high-resolution spectroscopy at each vortex, we used the numerical differentiation of $I$-$V$ curves.
All the data in this work were taken at the base temperature of the system where the effective electron temperature is estimated to be $\sim 85$~mK~\cite{Machida_RSI_2018}.

\subsection*{Additional information}
Correspondence and requests for materials should be addressed to T. M. or T. H.

\subsection*{Acknowledgments}
The authors thank C. -K. Chiu, A. Furusaki, P. A. Lee, D. -H. Lee, Y. Nagai, and T. T. Ong for valuable comments and C. J. Butler for critical reading.
This work was partly supported by CREST project JPMJCR16F2 from Japan Science and Technology Agency, a Grant-in-Aid for Scientific Research (A) (17H01141), and Japan–China Bilateral Joint Research Project by the Japan Society for the Promotion of Science (JSPS).

\subsection*{Author contributions}
T.M. carried out the experiments and the data analyses with assistance from Y. K. and T. H.
Y. S., S. P., T. T., S. T., and T. S. grew single crystals.
T. H. supervised the project.
T. M. and T. H. wrote the manuscript.

\subsection*{Competing financial interests}
The authors declare that they have no competing financial interests.

\end{document}